\documentclass[prb,aps,amssymb,twocolumn,showpacs,floatfix]{revtex4}

\usepackage{color}
\usepackage{epsfig}
\usepackage{graphicx}
\usepackage{dcolumn}
\usepackage{bm}
\usepackage{url}

\begin{document}

\title{Frequency- and electric-field-dependent conductivity of single-walled
carbon nanotube networks of varying density}
\author{Hua Xu \footnote[0$^{(a)}$]{$(a)$ email: umdxuhua@gmail.com}}
\author{Shixiong Zhang}
\author{Steven M. Anlage}
\affiliation{Center for Nanophysics and Advanced Materials,
Department of Physics, University of Maryland, College Park, MD
20742-4111}

\author{Liangbing Hu}
\author{George Gr\"{u}ner}
\affiliation{Department of Physics, University of California, Los
Angeles, CA 90095}

\begin{abstract}
We present measurements of the frequency and electric field
dependent conductivity of single walled carbon nanotube(SWCNT)
networks of various densities. The ac conductivity as a function of
frequency is consistent with the extended pair approximation model
and increases with frequency above an onset frequency $\omega_0$
which varies over seven decades with a range of film thickness from
sub-monolayer to 200 nm. The nonlinear electric field-dependent DC
conductivity shows strong dependence on film thickness as well.
Measurement of the electric field dependence of the resistance R(E)
allows for the determination of a length scale $L_{E}$ possibly
characterizing the distance between tube contacts, which is found to
systematically decrease with increasing film thickness. The onset
frequency $\omega_0$ of ac conductivity and the length scale $L_{E}$
of SWCNT networks are found to be correlated, and a physically
reasonable empirical formula relating them has been proposed. Such
studies will help the understanding of transport properties and
benefit the applications of this material system.
\end{abstract}

\maketitle

\section{Introduction}
Networks with randomly distributed single walled carbon nanotubes
(SWCNTs) are emerging as novel materials for various applications,
particularly as electronic materials.
\cite{Gruner,Ajayan,Baughmanr,CWZhou} This creates a need for the
accurate determination of their fundamental electrical properties.

SWCNT networks can be viewed as a two-dimensional network of
conducting one-dimensional rods.\cite{LHuNanolett} These rods are
either metallic or semiconducting with large dimension
ratio(length/diameter $\sim1000$), which leads to the following
interesting behavior.\cite{MSFuhrer,Dresse} Since the
nanotube-nanotube junction resistance is much larger than the
nanotube resistance itself,\cite{MSFuhrer,ABKaiserPRB,Stadermann}
such a network can be seen as a system with randomly distributed
barriers for electrical transport. For a sub-monolayer network, the
junction resistance will dominate the overall resistance and the
network resistance shows percolation behavior with nonlinear
thickness dependence.\cite{LHuNanolett,Stadermann} For SWCNT
networks with thickness in the tens of nanometers, e.g. with tens of
layers of SWCNTs, the conductance through the metallic tubes will
dominate the conduction, and leads to metallic behavior of the
overall resistance.\cite{Fischer,PPeit,SRoth}

The frequency-dependent and electric-field-dependent conductivity
have been investigated for individual
SWCNT\cite{Burke1,Burke2,MZhang} and some types of SWCNT
networks.\cite{ABKaiserCurr,APLshielding,PPeit,MSFuhrer1,MSFuhrer2}
Burke and co-workers measured the microwave conductivity of
individual SWCNTs and investigated their operation as a transistor
at 2.6 GHz.\cite{Burke1,Burke2} For SWCNT networks with 30 nm
thickness, our previous work finds the ac conductivity is
frequency-independent up to an onset frequency $\omega_0/2\pi$ of
about 10 GHz, beyond which it increases with an approximate power
law behavior.\cite{APLshielding} For thicker films with thickness in
the range of tens of micrometers, P. Peit \emph{et al.} found that
conductivities at DC and 10 GHz are almost the same.\cite{PPeit} M.
Fuhrer \emph{et al.} observed the nonlinearity of the electric
field-dependent conductivity of relatively thick nanotube networks,
and they claimed that the charge carriers can be localized by
disorder in the SWCNTs with an approximate length scale $l_{loc}$ of
1.65$\mu$m.\cite{MSFuhrer1,MSFuhrer2} V. Skakalova \emph{et al.}
studied the current-voltage(\emph{I-V}) characteristics of
individual nanotubes and nanotube networks of varying thickness, and
discussed the modulation of \emph{I-V} characteristics by a
gate-source voltage.\cite{SRoth} However there is a lack of
frequency and electric field dependent conductivity investigation
for systematic variation of network densities, and also there is no
correlation study between the onset frequency $\omega_0$ and the
length scale determined from nonlinear transport.

The frequency- and electric-field-dependent conductivity
investigation of different network densities is not only helpful to
build a comprehensive understanding of the electrical transport
properties of SWCNT networks, but is also important for the sake of
applications of these networks.\cite{APLshielding,SJkang,Zwu} For
example, when using SWCNT networks to construct high speed
transistors, or as transparent shielding materials, knowledge of
their frequency dependent electric transport properties are
required. Also use of SWCNT networks as inter-connects in circuits
requires understanding of their electric field-dependent properties.

In this paper, we investigated the frequency- and
electric-field-dependent conductivity for SWCNT networks with
systematically varying densities. We find that the onset frequency
$\omega_0$ extracted from the frequency dependence measurement
increases with the film thickness, while the length scale extracted
from the electric field dependent measurement decreases at the same
time. Using the measured $\omega_0$ and the extracted length scale
$L_{E}$ for different films, we developed an empirical relation
between the two transport measurement methods.

\section{Samples and experimental setup}

\begin{figure}[h]
\begin{center}
\epsfxsize=2.5in \epsfysize=3.8in \epsffile{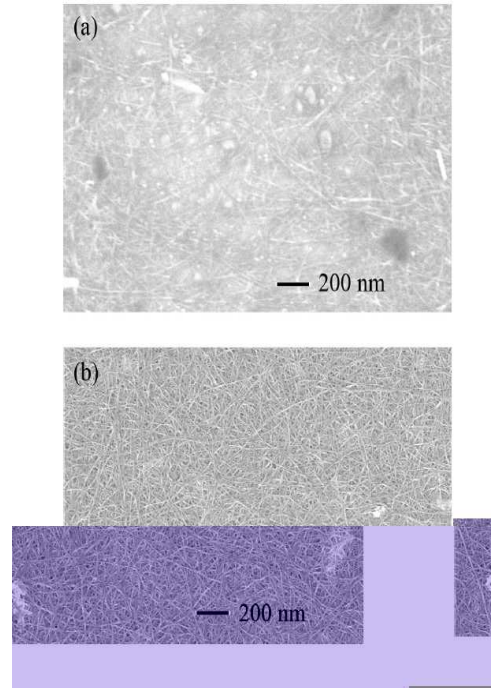}
\end{center}
\caption{SEM of two SWCNT network samples with different densities.
(a) sub-monolayer nanotube film; (b) nanotube film with thickness
about 20 nm.}\label{fig:Figuresample}
\end{figure}

SWCNT networks on polyethylene terephthalate (PET) substrate are
prepared by the spraying method.\cite{RHaddon} Arc-discharged
nanotube powder from Carbon Solution Inc. was dispersed in water
with sodium dodecyl sulphate surfactants. The tubes were purified
with nitric acid and left with a variety of functional
groups.\cite{RHaddon} After the nanotubes were sprayed on the
substrate, the samples were rinsed in water thoroughly to wash away
the surfactant. The network density, defined as the number of
nanotubes per unit area, was controlled by the solution
concentration(0.1 - 1 mg/mL) and the spraying time(10 - 60 seconds).
The nanotubes formed bundles with size of 4-7 nm and length of 2-4
$\mu$m.\cite{LHuAPL} The thicknesses of the films range from
submonolayer to 200 nm corresponding to room temperature sheet
conductance ranging from $10^{-6}$ to $10^{-2}$ S/Square. Fig.\
\ref{fig:Figuresample} shows SEM images of films with two different
coverage densities. It is convenient to characterize the films by
their sheet conductance rather than conductivity, because of the
inhomogeneous nature of the conducting pathways and thus the
difficulty in assigning a film thickness, particularly for
submonolayer films.

The Corbino reflectometry setup, which is ideal for broadband
conductivity measurements of resistive
materials,\cite{Jamesbooth,BoothRSI,Schwartz} was used to
investigate the conductivity of the SWCNT networks. The measurement
procedures are similar to our previous work.\cite{APLshielding} In
order to measure the conductivity in as broad a frequency range as
possible, two test instruments are used, the Agilent E8364B network
analyzer (covering 10 MHz to 50 GHz) and the Agilent 4396B network
analyzer (covering 100 kHz to 1.8 GHz), giving five and half decades
of frequency coverage. The electric field dependent conductivity was
measured with a standard 4-probe method in a well-shielded high
precision DC transport probe.\cite{Zhenglithesis}

\section{Frequency dependent conductivity}

\begin{figure}[h]
\begin{center}
\epsfxsize=3.2in \epsfysize=4.8in \epsffile{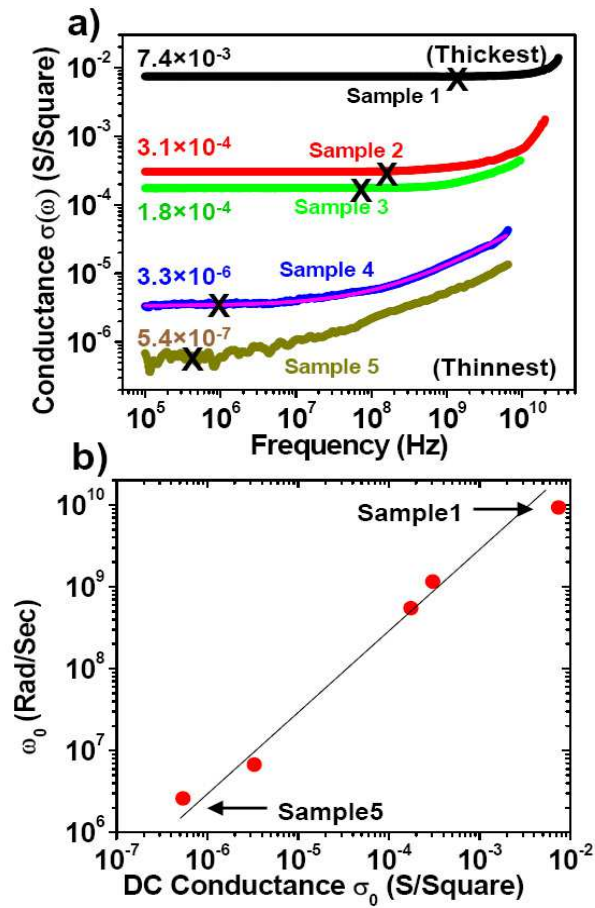}
\end{center}
\caption{(Color online) (a) Real parts of the ac conductance for
samples with different densities at room temperature. The black
crosses mark the fit onset frequency $\omega_0$ for each curve. The
magenta(grey) line is the typical fit result of the extended pair
approximation model for the sample 4 with
$\sigma_0=3.3\times10^{-6}$ S/Square, $\omega_0=1.1\times10^6$
Rad/Sec, $k=0.12$ and $s=0.52$. (b) The onset frequencies vs. dc
conductance of SWCNT networks. The solid line has slope 1.
}\label{fig:FigureCond}
\end{figure}

The data of conductance vs. frequency are shown in Fig.\
\ref{fig:FigureCond}(a). The films decrease in thickness from sample
1 (200nm) to sample 5 (sub-monolayer). For all the films, the real
parts of the conductance keep their dc value up to a characteristic
frequency and start to increase at higher frequency. This kind of
behavior has been widely observed for disordered
systems.\cite{JCDyre,PDutta,JCdyreRMP} Similar behavior has been
found for carbon nanotube polymer composites close to the
percolation threshold, in which the extended pair approximation
model was applied to describe the observed
phenomena.\cite{Kilbride,Mclachlan} The carbon nanotube networks in
Fig.\ \ref{fig:FigureCond}(a) have densities well above the
percolation threshold. However, since the junction resistances
between different tubes are much larger than the resistance of the
tubes themselves,\cite{MSFuhrer,ABKaiserCurr,Stadermann,
ABKaiserPRB} SWCNT networks above the percolation threshold can
still be seen as systems with randomly distributed barriers for
electrical transport. In this case the extended pair approximation
model can be used to describe the real part of the ac conductance:
\begin{eqnarray}\label{eq:EPAmodel}
\sigma = \sigma_0 (1 + k(\omega/\omega_0)^s)
\end{eqnarray}
where $\sigma_0$ is the dc conductance, $s\leq 1.0$, $k$ is a
constant and $\omega_0$ is the onset frequency.\cite{Kilbride}

The obtained frequency dependent conductance in Fig.\
\ref{fig:FigureCond}(a) fit well to the extended pair approximation
model. Fig.\ \ref{fig:FigureCond}(b) shows the relation between the
fit film onset frequency and the dc sheet conductance. The onset
frequency changes from $2\times10^6$ to $1\times10^{10}$ Rad/Sec as
the sheet conductance increases from $10^{-6}$ to $10^{-2}$
S/Square. The onset frequency of the ac conductance increases as the
dc conductance increases. The solid line Fig.\
\ref{fig:FigureCond}.(b) has slope one, implying a linear relation
between the onset frequency and their dc sheet conductance, or
$\omega_0 \sim \sigma_0$, over about 4 decades.

\begin{figure}[h]
\begin{center}
\epsfxsize=3.4in \epsfysize=2.8in \epsffile{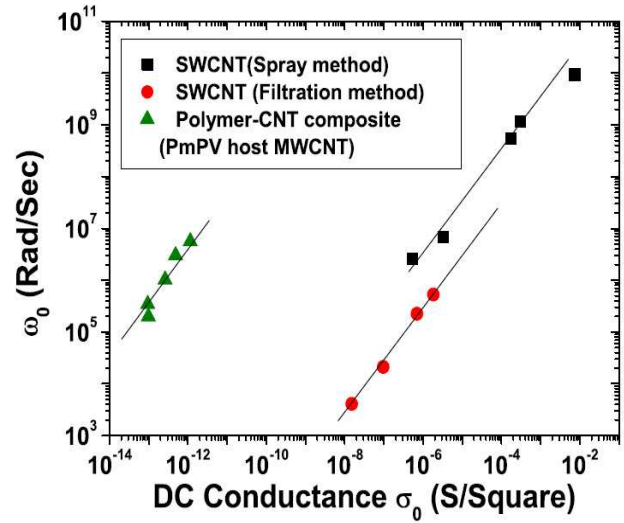}
\end{center}
\caption{(Color online) The onset frequency vs. dc conductance for
different SWCNT networks at room temperature. The three solid lines
all have slope one.}\label{fig:SWCNTFig4}
\end{figure}

We also measured the frequency-dependent conductance of ultra-thin
sub-monolayer nanotube networks, which are fabricated via the
filtration method.\cite{LHuNanolett} We choose Chloroform as the
solvent instead of water to avoid the washing steps which can easily
destroy the sub-monolayer film structure. The network density is
controlled by the concentration and volume of the solvent
used.\cite{LHuNanolett} The conductance of these films are measured
by a 4284A Precision LCR Meter, covering the frequency range from 20
Hz to 1 MHz. For these films just above the percolation threshold,
their frequency dependent conductance also fit well to the extended
pair approximation model. In Fig.\ \ref{fig:SWCNTFig4}, we plot the
onset frequency versus their dc sheet conductance. For comparison,
we also present results of Kilbride \emph{et al.},\cite{Kilbride}
for polymer-nanotube composites as well as the data from Fig.\
\ref{fig:FigureCond}(b) for our films prepared by the spray method.
The three solid lines in Fig.\ \ref{fig:SWCNTFig4} all have slope
one ($Rad/Sec \cdot S$). Here an interesting result is that although
the intercepts are very different, the slopes of the three different
samples are essentially the same. The difference in intercept is due
to the different types of nanotubes. Different carbon nanotube
sources have different bundle sizes and lengths, which causes the
conductance difference. The polymer-CNT composite has much smaller
DC conductance than the other two, due to the separation of SWCNTs
by polymer that leads to charge transfer barriers. Despite these
differences in detail, there is a universal relationship between
onset frequency and dc conductance.

\begin{figure}[h]
\begin{center}
\epsfxsize=3.4in \epsfysize=2.8in \epsffile{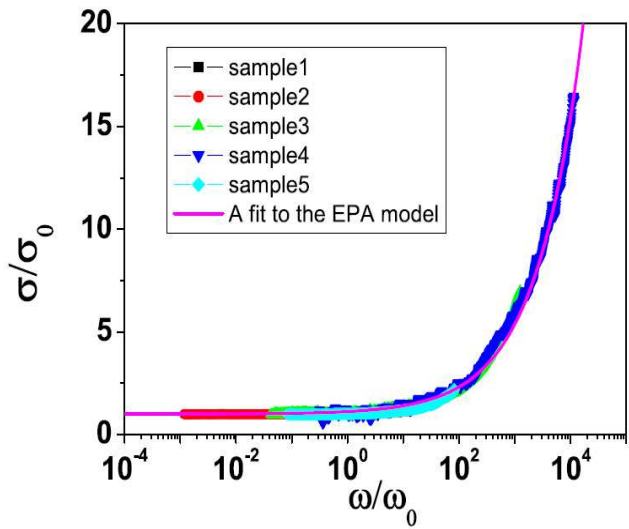}
\end{center}
\caption{(Color online) Master scaling curve showing the ac
conductance for SWCNT network samples(spray method) with different
densities at room temperature. The red(grey) solid line is a fit to
the extended pair approximation model $k=0.12$ and
$s=0.52$.}\label{fig:SWCNTscaling}
\end{figure}

The linear proportionality $\omega_0\sim\sigma_0$ is one of the
characteristic features of ac conductance of disordered
solids.\cite{JCdyreRMP} For disordered systems, the universality of
ac conductance has been studied by many researchers since its
discovery in the 1950's, and is usually referred to as Taylor-Isard
scaling. \cite{JCDyre,PDutta,JCdyreRMP} This inspires us to
investigate the scaling behavior of the conductance and frequency
for SWCNT networks. Similar to Kilbride \emph{et al.}'s
work\cite{Kilbride}, taking the dc conductance $\sigma_0$ and the
onset frequency $\omega_0$ as scaling parameters, we plotted the ac
conductance data for all samples in Fig.\ \ref{fig:SWCNTscaling}.
The data show a scaling behavior with all data sets falling on the
same master curve described by $k=0.12$ and $s=0.52$.

In order to explain the observed onset frequency change with the
SWCNT volume fraction for polymer-nanotube composites close to
percolation threshold, Kilbride \emph{et al.} defined a
characteristic length scale, the "correlation length"
$\xi$.\cite{Kilbride} For our samples, which are well above the
percolation threshold, we can similarly define the correlation
length $\xi$ as the distance between connections in the sample,
i.e., the distance between junctions of the multiply-connected SWCNT
network. When one measures the conductance at a given frequency,
there is a typical probed length scale. At low frequencies
(corresponding to long time periods), the carriers travel long
distances in one-half ac cycle and the experiment investigates
longer length scales.  At high frequencies, the carriers travel
short distances in one half ac-cycle and the probed length scale is
shorter. In the absence of an applied DC electric field, and
assuming a random walk, the probed length scale
$L_{\omega}\propto\omega^{-1/2}$.\cite{Kilbride}

For low frequency, $L_{\omega}>\xi$, the probed length scale spans
multiple junctions of the SWCNT network. The junction resistances
between different tubes are much larger than the resistances of the
tubes themselves.
\cite{MSFuhrer,ABKaiserCurr,Stadermann,ABKaiserPRB} Hence the
conductance is small and equal to the dc conductance. As the
frequency increases, $L_{\omega}$ becomes smaller than $\xi$,
$L_{\omega}<\xi$, and the junction resistances have less
contribution to the total resistance. The intrinsic properties of
the SWCNT then dominate the measurement. Therefore the observed
conductance increases as the frequency increases. The transition
frequency is expected to be that where carriers scan an average
distance of order of the correlation length. \cite{Kilbride}

The SWCNT film can be treated as a random walk network, in which the
probed length scale at a given frequency goes as
$L_{\omega}\sim\omega^{-1/2}$. Then with the onset frequency
$\omega_0$ of samples we are able to estimate their correlation
length $\xi\simeq L_{\omega_0}\propto \sqrt{\frac{1}{\omega_0}}$. As
the density of the SWCNT networks increases, there are more
junctions and the average distance between connections, the
correlation length $\xi$, becomes smaller and thus the onset
frequency $\omega_0$ increases. This is consistent with the behavior
shown by the data in Fig.\ \ref{fig:FigureCond}.

\section{Electric field dependent conductivity}

From the measured frequency dependent conductivity, we observed that
the correlation length $\xi$ varies as the thickness of the film
increases. Due to the large resistance of junctions between
different tubes, the carriers in the SWCNT networks are easier to
move on small length scales than on larger length scales. This
phenomenon inspires us to investigate whether the transport
properties are nonlinear as a result. According to the works of M.
S. Fuhrer \emph{et al.}, the localization behavior of a SWCNT
network can be studied by measuring the electric field dependent
nonlinearity of the conductivity. \cite{MSFuhrer1,MSFuhrer2}
Generally, increasing both the temperature and the electric field
reduces the effects of localization. Hence a system typically
displays a characteristic electric field at which nonlinear
conductivity begins to appear. Through measurement of the
characteristic field $E_c$ one can determine a length scale of the
system:
\begin{eqnarray}\label{locallengthscale}
L_{E}= \frac{k_BT}{e E_c}
\end{eqnarray}
where $L_{E}$ corresponds to the size of the regions with good
conductivity, and $eEL_{E}$ is the energy gained by a carrier in one
conducting region on average.\cite{KMortensen,NApsley}. For our
SWCNT films $L_E$ depends on the average distance between
connections and thus should vary with sample thickness. For
relatively thick films($\sim 10 \mu m$) with low sheet resistance,
Fuhrer \emph{et al.} found a temperature-independent length scale
$l_{loc}\sim1.65\mu$m in a SWCNT film.\cite{MSFuhrer1,MSFuhrer2}

Here we investigated the electric field-dependent conductance for a
different class of films. The studied samples were cut into
rectangular pieces with length $\sim8$mm and width $\sim3$mm. Gold
contact leads, which completely cover the width of the samples, were
deposited on the surfaces of the samples to form a standard four
probe pattern. The measured SWCNT networks have dimensions of
$\sim3.5$mm$\times$3mm, between the two voltage leads. Because SWCNT
films have a large in-plane thermal conductivity, \cite{Hone,Ouyang}
and the probe used to perform the measurement was specially designed
to reduce heating effects\cite{Zhenglithesis} we believe that the
heating effect can be ignored in our measurement range. This was
confirmed by observing the voltage response of pulsed current input
of several samples in the time domain with an oscilloscope.

\begin{figure}[h]
\begin{center}
\epsfxsize=3.2in \epsfysize=4.8in  \epsffile{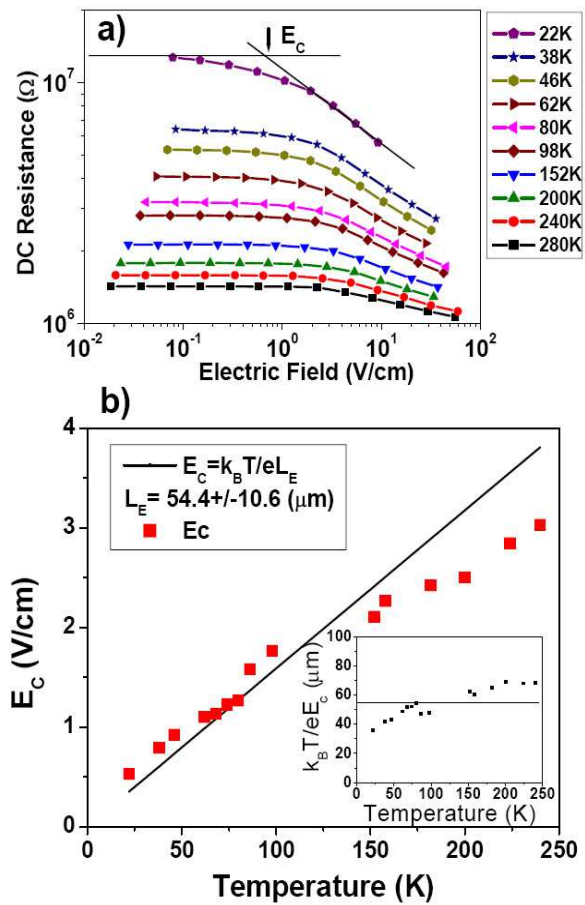}
\end{center}
\caption{(color online) (a) DC resistance vs electric field for a
submonolayer(sample 5) SWCNT network at different temperatures. (b)
$E_c$ vs temperature and extracted length scale for this
sample.}\label{fig:SWCNTFig5}
\end{figure}

The DC resistance vs. electric field of a submonolayer film(sample
5) at different temperatures is shown in Fig.\
\ref{fig:SWCNTFig5}(a). We can see clearly that there is nonlinear
behavior even at room temperature. As sketched in Fig.\
\ref{fig:SWCNTFig5}(a), we extracted $E_c$ for each temperature,
which is the characteristic electric field at which nonlinear
conductance begins to appear.\cite{MSFuhrer1} Fig.\
\ref{fig:SWCNTFig5}(b) shows the extracted $E_c$ vs temperature for
this sample. As temperature increases, $E_c$ also increases.
Roughly, $E_c$ and temperature have a linear relationship and
satisfy the equation $E_c = k_BT/eL_{E}$, with $L_{E}\approx54\mu
m$. Note that this is much larger than the $l_{loc}$ obtained in
Ref.\cite{MSFuhrer1,MSFuhrer2}. This suggests that a different
mechanism for hindering transport is active here, namely poor
contacts between bundles of SWCNTs.

\begin{figure}[h]
\begin{center}
\epsfxsize=3in \epsfysize=2.6in \epsffile{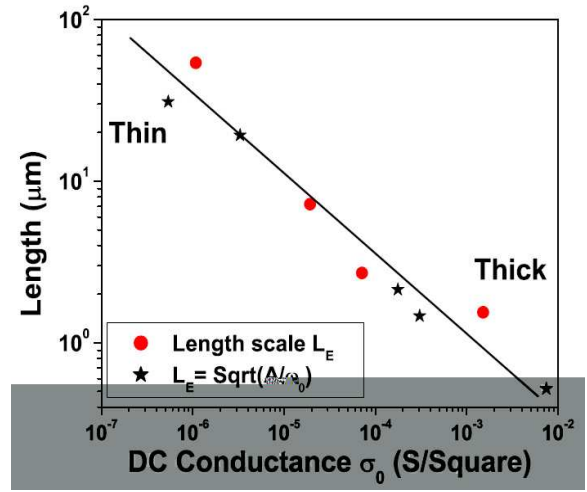}
\end{center}
\caption{(Color online) Comparison of the length scale $L_E$
obtained from electric field dependent DC conductance data(red dot),
and the estimated length scale from frequency dependent conductance
measurement $\sqrt{A/\omega_0}$ (black star) of SWCNT networks of
various densities. The line has slope -1/2 in the log-log plot.
}\label{fig:SWCNTFig6}
\end{figure}

To investigate the dependence of $L_E$ on the film thickness, we
measured the electric field-dependent conductance for samples with
various thicknesses. We found that the nonlinear behavior strongly
depends on the thickness of the film. At low temperature, all the
samples show nonlinear behavior of resistance on electric field, and
the critical field $E_c$ is larger for thicker samples and smaller
for thin samples. At room temperature, all the other samples are
purely ohmic(linear) in electric fields up to 100 V/cm, except for
the thinnest sample, sample 5, which showed electric field-dependent
conductance at all temperatures. Through the critical field $E_c$,
we extracted the $L_E$ for different density films, shown in Fig.\
\ref{fig:SWCNTFig6} as red-dots. Clearly $L_E$ depends on the
density of the SWCNT networks. For the thicker film with larger
SWCNT density, which has larger sheet conductance, the $L_E$ is
smaller. The lower density films with larger average distance
between the junctions has larger $L_E$.

Although the localization behavior may be affected by the properties
of individual SWCNTs \cite{MSFuhrer1,MSFuhrer2}, the above result
shows that the observed behavior is more likely a consequence of
poor electric conduction through inter-bundle junctions. As the
thickness and density increase, there are more and more connections,
which decreases the length scale $L_E$, as shown in Fig.\
\ref{fig:SWCNTFig6}.

When the film thickness increases to a large value, the density of
the SWCNTs and connections of the film will finally saturate and
stop increasing. For those very thick SWCNT films (also called
mats), their correlation length $\xi$, and also the length scale
$L_E$, as well as the onset frequency, will be independent of the
thickness, but depend on the properties of individual bundles, which
include both the disorder of the individual nanotube and also the
geometry of the bundles. Hence the difference of SWCNT sources and
purities can change the onset frequency. We believe this is the
reason that our previous work found that the ac conductivity follows
the extended pair approximation model with an onset frequency
$\omega_0$ about 10 GHz, while Peit \emph{et al.} found that SWCNT
networks have the same conductivities at DC and 10 GHz for SWCNT
networks with thickness in the range of tens of
micrometers.\cite{APLshielding,PPeit}

\section{Discussion}

The frequency dependent conductivity shows that the onset frequency
$\omega_0$ increases with the sample thickness and gives an
estimation of the correlation length $\xi$ of the sample, $\xi\sim
L_{\omega_0}\propto\sqrt{\frac{1}{\omega_0}}$. $\xi$ and
$L_{\omega_0}$ decrease as the sample thickness increases. The DC
nonlinear conductivity measurement shows that the length scale $L_E$
of the samples also decreases with increasing sample thickness.
Qualitatively $L_{\omega_0}$ and $L_E$ have the same dependence on
the sample thickness. Based on the above consideration, it is
reasonable to expect some correlation between the onset frequency
$\omega_0$ and the length scale $L_E$, and we proposed the following
relation,
\begin{eqnarray} \label{eq:emperequation}
L_{E} = \sqrt{\frac{A}{\omega_0}},
\end{eqnarray}
where $A$ is a fitting parameter, which can be obtain from our
experiment data, $A\sim2.5\pm0.5\times10^{-3} m^2/s$. Fig.\
\ref{fig:SWCNTFig6} shows the comparison of the nonlinear transport
length scale $L_E$ measured directly through electric-field
dependent DC conductance measurements(red-dot) and the right-hand
side of Eq.\ \ref{eq:emperequation} from the frequency dependent
measurement(black-star). They are consistent with each other. In
this way we obtain an empirical relation between the
frequency-dependent and the electric field-dependent conductivity.
The parameter $A$ should depend on the diffusion properties of
carriers in the SWCNT films.\cite{Minnhagen} Writing
$L_E=\sqrt{4Dt_*}$ , where D is a diffusion constant and $t_*$ is
the corresponding time scale, we take $D=A/8\pi$. Assuming the
Einstein relation $\mu = \frac{De}{k_BT}$, we can estimate the
mobility of carriers on the SWCNT network, which is on the order of
$10^{-2} m^2/V\cdot sec$. This is much smaller than that of the
individual single walled carbon nanotube, which can be on the order
of $10^0 - 10^1 m^2/V\cdot sec$ at room
temperature.\cite{FuhrerMobility1,FuhrerMobility2} In the SWCNT
networks, due to the randomly distributed barriers for electrical
transport, the mobility is expected to be much smaller than the
SWCNT itself. The carbon nanotube film-based transistors have
reported mobilities on the order of $10^{-4}-10^{-1} m^2/V\cdot
sec$,\cite{Artukovic,Weixue,SJkang} which is consistent with our
estimation. To get a deeper physical understanding of these
networks, more investigation and theoretical work are needed. For
example, measuring the temperature dependence of the
frequency-dependent conductivity for SWCNT networks with various
densities would be one approach.\cite{PPeit,APLshielding}

\section{Conclusion}
To conclude, we systematically studied the frequency and electric
field dependent conductivity of SWCNT films of various thicknesses.
We found the poor interbundle junctions affect the transport
behavior of the carriers in the SWNCT films, which causes strong
thickness dependence of characteristic length scales. A thinner film
has larger correlation length $\xi$ and length scale $L_E$, so it
has a smaller onset frequency $\omega_0$, whereas a thicker film
with smaller correlation length $\xi$ and length scale $L_E$, has a
larger onset frequency $\omega_0$. An approximate empirical formula
relating the onset frequency $\omega_0$ and nonlinear transport
length scale establishes contact between the frequency and the
electric field dependent conductivity, which helps us to
understand the electric transport properties of the SWCNT films.\\

The authors would like to thank M. S. Fuhrer, C. J. Lobb, Weiqiang
Yu and Bing Liang for their help and discussions on this work. This
work has been supported by Center for Nanophysics and Advanced
Materials of the University of Maryland and the National Science
Foundation Grant Nos. (DMR-0404029), (DMR-0322844), and
(DMR-0302596).

\end{document}